\input{aipcheck}

\documentclass[
    ,final            
  ]
  {aipproc}

\layoutstyle{6x9}

\begin{document}

\title{Azimuthal Asymmetries: \\ Access to Novel Structure 
Functions~\footnote{Presented at the 15th International Spin Physics 
Symposium (Spin 2002), Brookhaven National Laboratory, September 9-14, 
2002.}}

\author{K. A. Oganessyan}{
 address={INFN-Laboratori Nazionali di Frascati, Enrico Fermi 40,
  I-00044 Frascati, Italy}
  ,altaddress={DESY, Notkestrasse 85, 22603 Hamburg, Germany}
}

\author{L. S. Asilyan}{
  ,address={INFN-Laboratori Nazionali di Frascati, Enrico Fermi 40,
  I-00044 Frascati, Italy}
}

\author{E. De Sanctis}{
  ,address={INFN-Laboratori Nazionali di Frascati, Enrico Fermi 40,
  I-00044 Frascati, Italy}
}

\author{V. Muccifora}{
  ,address={INFN-Laboratori Nazionali di Frascati, Enrico Fermi 40,
  I-00044 Frascati, Italy}
}

\begin{abstract}
One of the most interesting consequence of non-zero intrinsic 
transverse momentum of partons in the nucleon is the nontrivial 
azimuthal dependence of the cross section of hard scattering processes. 
Many of the observable asymmetries contain unknown functions which 
provide essential information on the quark  and gluon structure.  
Several  of them have  been studied in the last few years; we discuss 
their qualitative and quantitative features in semi-inclusive DIS.
\end{abstract}

\maketitle

\subsection{Introduction}

The study of the structure of hadrons, bound states of
quarks and gluons, in the context of quantum chromodynamics 
(QCD), is one of the challenges of elementary particle 
physics requiring new nonperturbative approaches in the 
field theory. 

This talk focuses on a discussion of correlations between 
the spins of hadron or quark and the momentum of the quark 
with respect of that of the hadron in polarized hard scattering 
processes. Signatures of these correlations appear in 
high-energy scattering processes as correlations between 
the azimuthal angles of the (transverse) spin and the 
(transverse) momentum vectors. 
 
In particular, we discuss single-spin (transversely polarized 
target or longitudinally polarized beam) and double-spin (longitudinally 
polarized beam and transversely polarized target) azimuthal asymmetries 
in single hadron electroproduction in DIS. 

\begin{figure}
  \includegraphics[height=.3\textheight]{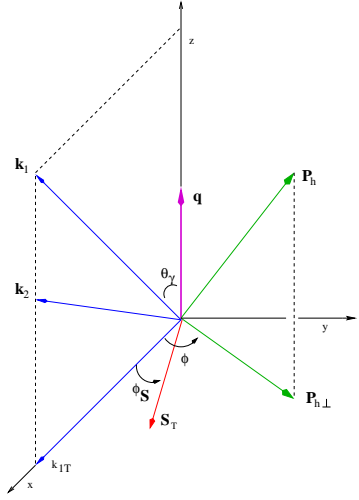}
  \caption{The kinematics of semi-inclusive DIS: $k_1$ ($k_2$) 
is the 4-momentum of the incoming (outgoing) charged lepton, 
$Q^2=-q^2$, where $q=k_1-k_2$, is the 4-momentum of the virtual 
photon. The momentum $P$ ($P_h$) is the momentum of the target 
(observed) hadron. The scaling variables are $x=Q^2/2(P\cdot q)$, 
$y=(P\cdot q)/(P\cdot k_1)$, and $z=(P\cdot P_h)/(P\cdot q)$.  
The momentum $k_{1T}$ ($P_{h\perp}$) is the incoming lepton (observed hadron) 
momentum component perpendicular to the virtual photon momentum direction, 
and $\phi$ is the azimuthal angle between $P_{h\perp}$ and $k_{1T}$,  
$S_T$ is the target spin vector and $\phi_S$ is its azimuthal angle. }
\end{figure}


The general form of the factorized cross sections of hard scattering 
process is written as~\cite{QS}
\begin{equation}
d\sigma = H^0 \otimes f_2 \otimes f^{'}_2 
+ {1 \over Q^n} H^1 \otimes f_2 \otimes f^{'}_{2+n} + {\cal O}
{ \left( {1 \over Q^{n+1}} \right) }, 
\label{FCS}
\end{equation}
with $n=1$ and $2$ for the polarized and unpolarized case, respectively. 
The perturbatively calculable coefficient functions are denoted $H^0, 
H^1$ and are convoluted with the nonperturbative soft parts $f_t, f'_t$, 
where $t$ denotes the twist. The cross section is related to  
helicity-dependent amplitudes 
${\cal{A}}_{\Lambda\lambda,\Lambda'\lambda'}$~\cite{JAF1}, 
which describes a process where a target of helicity $\Lambda$ 
emits a parton of helicity $\lambda$, and the scattered parton with 
helicity $\lambda'$ is reabsorbed by a hadron of helicity $\Lambda'$. 
In the spin-1/2 case, due to helicity conservation, parity and 
time invariance, for each twist assignment there are three 
independent 
helicity amplitudes. The familiar distribution functions~\footnote{The 
asterisk indicates bad component of quark/gluon field.} 
$f_1 \Leftrightarrow ({\cal A}_{+\,+,+\,+}+{\cal A}_{+\,-,+\,-})$, 
$g_1 \Leftrightarrow ({\cal A}_{+\,+,+\,+}-{\cal A}_{+\,-,+\,-})$, 
and $g_T \Leftrightarrow ({\cal A}_{+\,{+}^{*},+\,-})$ can be measured 
in inclusive DIS. The chiral-odd nature of  
$h_1 \Leftrightarrow ({\cal A}_{+\,-,-\,+})$, 
$h_L \Leftrightarrow ({\cal A}_{+\,{+}^{*},+\,+}-{\cal A}_{+\,-,+\,-})$, 
and 
$e \Leftrightarrow ({\cal A}_{+\,{+}^{*},+\,{+}}+{\cal A}_{+\,-,+\,-})$
makes their experimental determination difficult. Since electroweak 
and strong interactions conserve chirality, these functions cannot occur 
alone, but have to be accompanied by a second chiral odd quantity. For 
this reason, up to now no experimental information on these functions 
is available.

\subsection{Transverse target single-spin azimuthal asymmetry}

To access the transversity distribution function, $h_1$ (also commonly 
denoted $\delta q$), in semi-inclusive DIS off transversely polarized 
nucleons (see the relevant kinematics in Fig.1), one can measure the 
azimuthal angular dependences in the production of leading spin-0 
or (on average) unpolarized hadrons from transversely polarized quarks 
with nonzero transverse momentum. This production is described 
by the intrinsic transverse momentum dependent fragmentation function 
$H_1^{\perp}(z)$, which is chiral-odd and also  T-odd, i.e., 
non-vanishing only due to final state interactions~\cite{COL}. 
Due to its chiral-odd structure it is a natural partner to isolate 
chiral-odd distribution functions, such as $h_1$, $h_L$, and $e$. 
It is worthy to note that a clean separation of current 
and target fragmentation effects in the data is 
required~\cite{BERG,MULD}.      

The observable moment is defined as the appropriately weighted integral 
over $\phi$ of the cross section asymmetry~\footnote{The first and second 
subscripts indicate the polarizations of beam and target,  
respectively. We use $U$ for unpolarized, $L$ for longitudinally 
polarized and $T$ for transversely polarized particles.}:

\begin{equation}
A^{\sin(\phi+\phi_S)}_{UT} \equiv 
\frac{\int d\phi \sin(\phi+\phi_S) \left[ d\sigma^{\Uparrow}
-d\sigma^{\Downarrow} \right]} 
{{1 \over 2} \int d\phi \left [ d\sigma^{\Uparrow}+
d\sigma^{\Downarrow} \right]}, 
\label{AS}
\end{equation}
where $\Uparrow (\Downarrow)$ denotes the up (down) 
transverse polarization of the target in the virtual-photon frame. 
This asymmetry is given by~\cite{COL,AK,TM}: 
\begin{equation}
\label{UA}
A^{\sin(\phi+\phi_S)}_{UT} \propto \frac{h^u_1(x)}{f^u_1(x)} 
\cdot \frac{H_1^{\perp(1)u}(z)}{D^u_1(z)}.  
\end{equation}

An indication of a non-zero $H^{\perp (1)}_1(z)$ comes from 
the single spin asymmetry measured for pions produced in 
semi-inclusive DIS of leptons off a longitudinally polarized 
target at HERMES~\cite{HERM}. 

  \begin{figure}
  \includegraphics[height=.2\textheight]{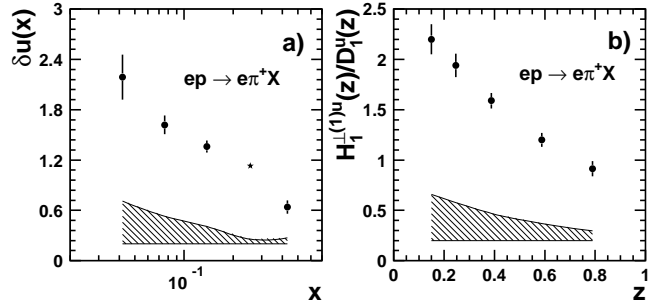}
  \caption{{\bf a)} The transversity distribution $\delta u(x)\equiv 
   h^u_1(x)$ and {\bf b)} the ratio of the fragmentation functions 
   $H^{\perp (1) u}_1(z)$ over $D_1(z)$ as they would be measured at 
   HERMES~\cite{KNO,HIN}. The hatched bands show projected systematic 
   uncertainties. } 
\end{figure}

A larger asymmetry with a transversely polarized target is 
expected~\cite{OBDN,BAC}. However, the existence of the competing 
mechanism, which, due to asymmetric distribution of quarks 
transverse momenta in a hadron, also gives a transverse spin 
asymmetry at leading twist~\cite{BROD,COLN}, turns the transversity 
measurement challenging. For distinguishing the different mechanisms 
and for its complete description ($x$-dependence at large $x$, 
$Q^2$-evolution, etc.) results from different scattering processes 
and different kinematics are required. New data from HERMES and 
COMPASS measurements on transversely polarized targets will give 
possibility to measure the transversity~\cite{KNO}. Fig.2 show the 
expected accuracies for the reconstruction of $h^u_1(x)$ and 
$H^{\perp (1) u}_1(z)/D_1(z)$ at HERMES with transversely polarized 
proton target~\cite{KNO,HIN}. 
In addition, analysis of data from $e^{+}e^{-} \to \pi^{+}\pi^{-}X$ 
process, which expected to show a similar azimuthal correlations is 
under way at the BELLE B-factory 
at KEK~\cite{BELLE}. 

\subsection{Beam single-spin azimuthal asymmetry}

 \begin{figure}
\includegraphics[height=0.2\textheight]{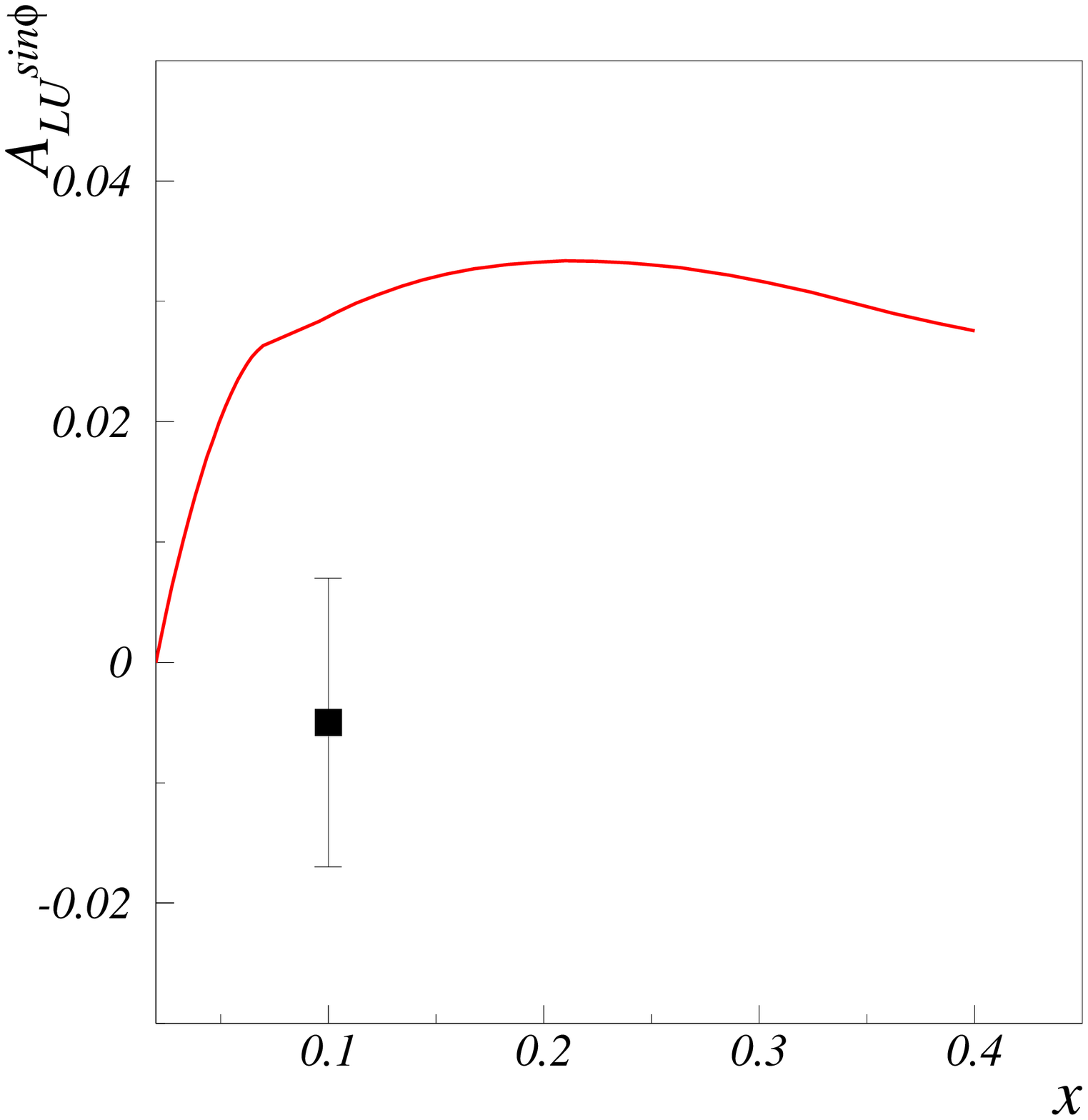}
\includegraphics[height=0.2\textheight]{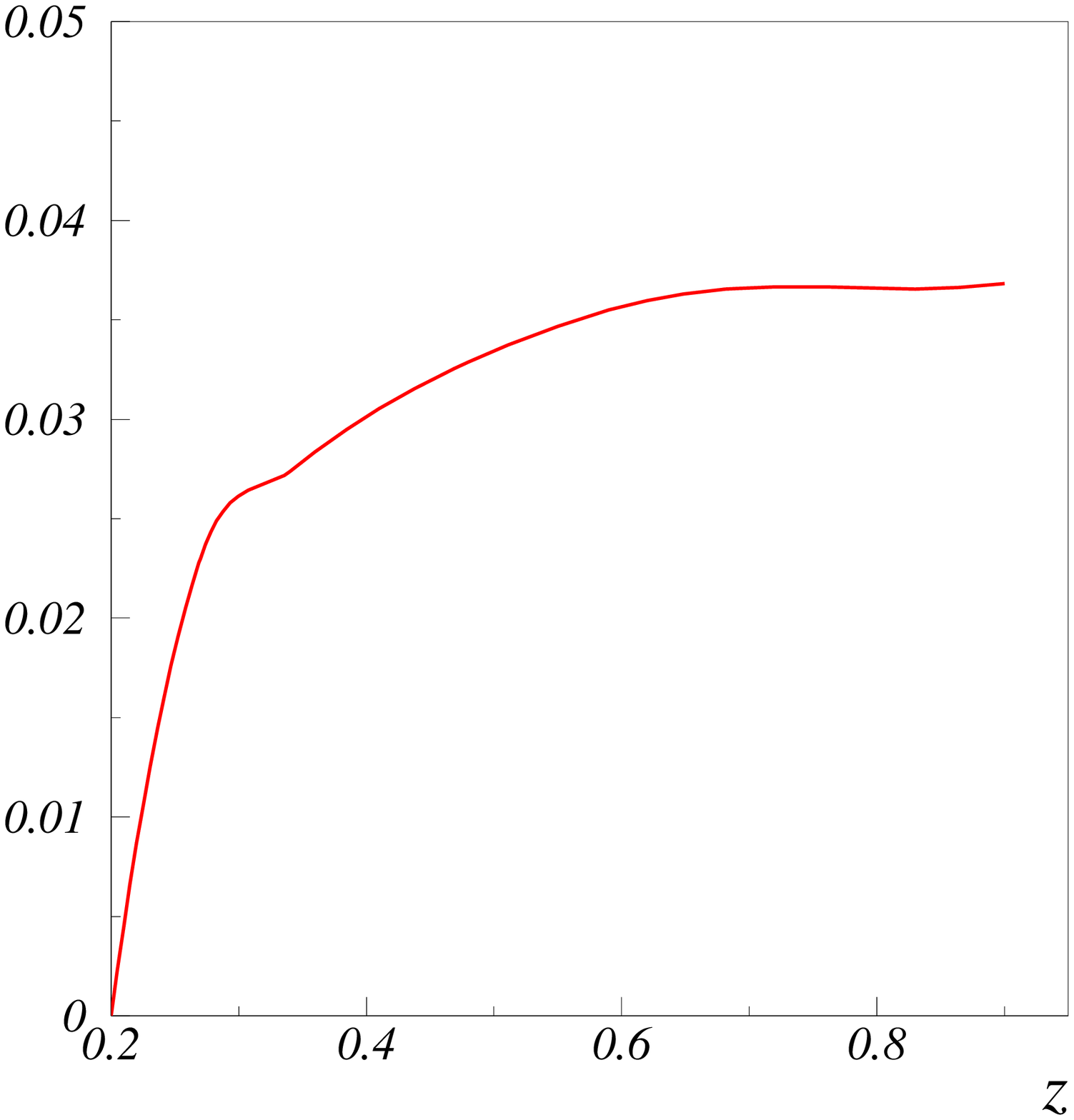} 
\end{figure}  
\begin{figure}
  \includegraphics[height=0.2\textheight]{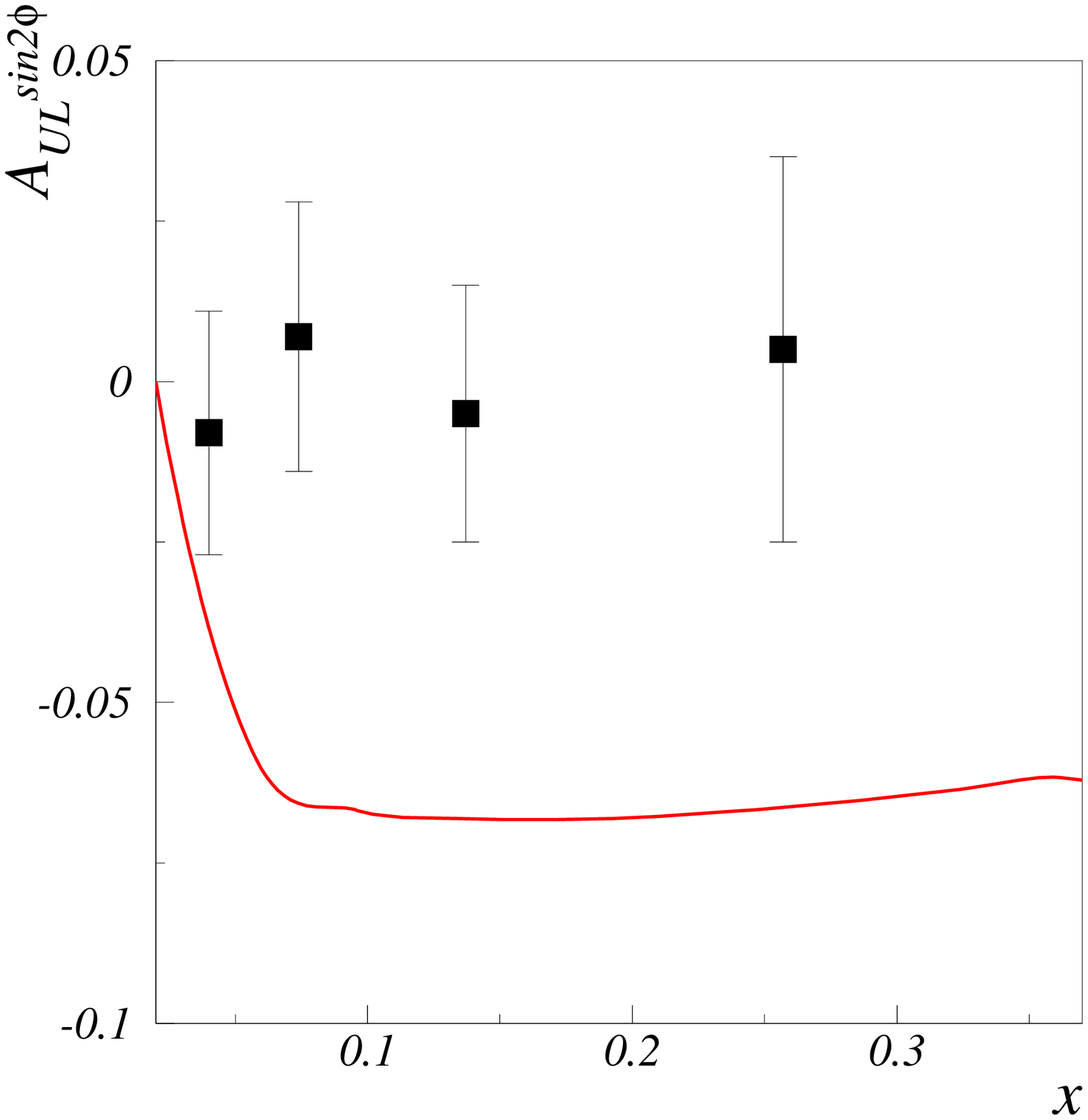}
  \includegraphics[height=0.2\textheight]{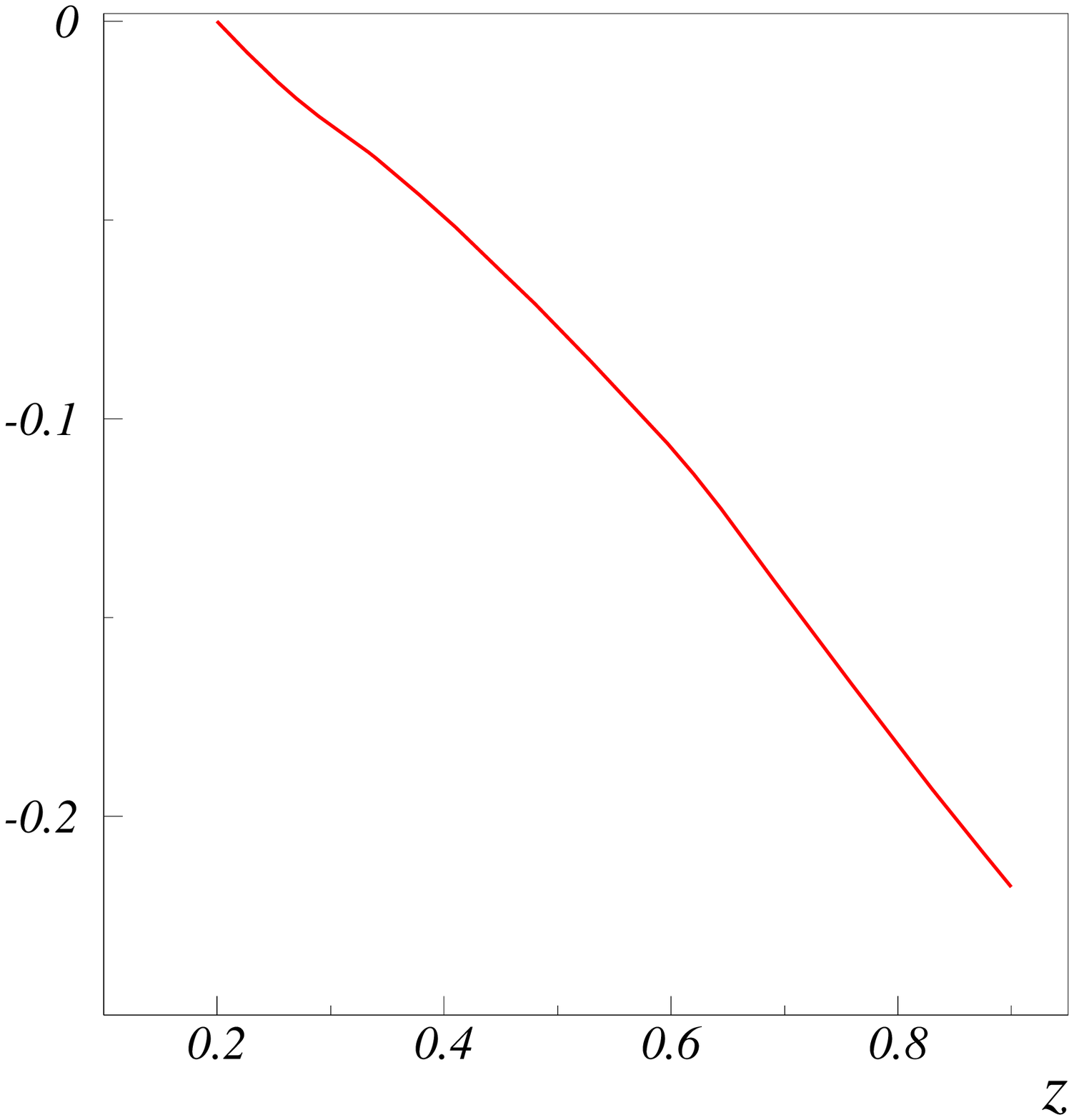}
  \caption{The single beam spin $A^{sin\phi}_{LU}$ and the 
single target spin $A^{sin2\phi}_{UL}$ asymmetries 
for $\pi^+$ production as a function of $x$ and $z$. Data are 
from  HERMES experiment~\cite{HERM}. Error bars show the statistical 
and the systematical uncertainties.} 
\end{figure}
  
At order $1/Q$ a $\sin\phi$ asymmetry was predicted~\cite{LM} for 
longitudinally polarized beam and unpolarized target. It probes the 
interaction-dependent distribution function $\tilde{e}(x)$ 
($\tilde{e}(x) = e(x) - {m \over M} {f_1(x) \over x}$)~\cite{TM,LM} 
in combination with the above mentioned fragmentation function 
$H^{\perp}_1$. It is important to notice  
that this asymmetry is related to the left-right asymmetry in the hadron 
momentum distribution with respect to the electron scattering plane, 
\begin{equation}
A_{LU} = \frac{\int_0^\pi d\phi d\sigma 
- \int_\pi^{2\pi} d\phi d\sigma}{\int_0^\pi d\phi d\sigma 
+ \int_\pi^{2\pi} d\phi d\sigma}, 
\label{A2}     
\end{equation}  
which is $2/\pi$ times $A^{\sin\phi}_{LU}$,
\begin{equation}
A^{\sin\phi}_{LU} = \frac{\int d\phi \sin \phi [d\sigma^{\rightarrow} 
- d\sigma^{\leftarrow}]}
{{1 \over 2} \int d\phi [d\sigma^{\rightarrow}
+d\sigma^{\leftarrow }]} \quad \propto \frac{\sum_a e^2_a  
\tilde{e}^a(x) H_1^{\perp (1) a}(z)}
{\sum_a e^2_a f_1^a(x) D_1^{a}(z)}. 
\label{LU}
\end{equation}
 
To evaluate this asymmetry as well as the single target-spin $\sin2\phi$ 
asymmetry, $A^{sin2\phi}_{UL}$~\cite{TM,DNO}, we use the MIT bag 
model~\cite{JAF2} as input.
 
For the weighted T-odd fragmentation function, $H^{\perp(1)}_1(z)$, 
we use the Collins ansatz~\cite{COL} for the analyzing power with 
the factor $\eta=1.6$~\cite{OBDN}. The $x$- and $z$-dependences of the 
$A^{sin\phi}_{LU}$ and $A^{sin2\phi}_{UL}$ for $\pi^+$ are shown 
in Fig.3. The asymmetry $A^{sin\phi}_{LU}$ amounts to about 3{\%}, while 
$A^{sin2\phi}_{UL}$ is larger (about 6{\%}). This is in disagreement 
with the asymmetries measured by HERMES~\cite{HERM}, which, as shown 
in the figure, are consistent with zero within the errors. This clearly 
indicate the necessity of a serious improvement of MIT bag model.     

\subsection{Double-spin azimuthal asymmetry}

The double spin azimuthal asymmetry is related to the $g_T(x)($$g_2(x))$ 
distribution function.  

Accounting for transverse momenta of the quarks, a longitudinal quark spin
asymmetry exists in a transversely polarized nucleon target.
The relevant leading twist distribution $g_{1T}(x, p_T^2)$ can be
determined from the measurement of this asymmetry in semi-inclusive 
DIS~\cite{TM,KM,OMD} in the case of longitudinally polarized beam 
and transversely polarized target.

The observable $\cos(\phi-\phi_S)$ moment in the 
cross section is defined in the following way
\begin{equation}
A^{\cos(\phi-\phi_S)}_{LT}= 
\frac{\int d\phi \cos (\phi - \phi_S) \cdot [ \sigma^{\leftarrow \Uparrow}+\sigma^{\rightarrow \Downarrow}
-\sigma^{\leftarrow \Downarrow}-\sigma^{\rightarrow \Uparrow}]} 
{\int d\phi \cdot [\sigma^{\leftarrow \Uparrow}
+\sigma^{\rightarrow \Downarrow} 
+\sigma^{\leftarrow \Downarrow}+\sigma^{\rightarrow \Uparrow}]}
\label{ASMY}
\end{equation}
\begin{equation}
A^{\cos(\phi-\phi_S)}_{LT} \propto \frac{\sum_a e^2_a g_{1T}^{(1)}(x)\, z\, D_1(z)}
{\sum_a e^2_a f_1(x) D_1(z)}. 
\label{ASMY1}
\end{equation}

  \begin{figure}
  \includegraphics[height=0.2\textheight]{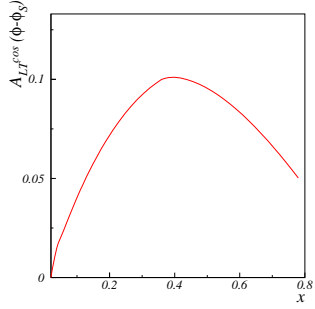}
  \caption{The double-spin asymmetry $A^{cos(\phi-\phi_S)}_
{LT}$ for $\pi^+$ production as a function of $x$.}
\end{figure}     


The $x$-dependence of $A^{cos(\phi-\phi_S)}_{LT}$ calculated for 
$\pi^+$ production is shown in Fig.4~\cite{KM,OMD}: as seen it is a 
sizable asymmetry that  may provide an alternative way to measure 
$g_2(x)$.    

\subsection{Conclusion}

We have discussed the qualitative and quantitative features 
of nontrivial azimuthal dependence of the cross section of the 
semi-inclusive DIS which provided essential information on nucleon 
spin structure.

\end{document}